\documentclass[runningheads]{llncs}
\usepackage{graphicx}
\usepackage[utf8]{inputenc}
\usepackage{xurl}
\usepackage{cite}
\usepackage{multirow}
\usepackage{csquotes}
\usepackage{array}
\usepackage{amssymb}
\usepackage{amsmath}

\begin{document}
\title{Recommending best course of treatment based on similarities of prognostic markers\thanks{All authors contributed equally.}}
\titlerunning{Recommending treatment based on prognostic markers}

\author{Sudhanshu\and
Narinder Singh Punn\and
Sanjay Kumar Sonbhadra\and
Sonali Agarwal}

\authorrunning{Sudhanshu et al.}

\institute{Indian Institute of Information Technology Allahabad, India\\
\email{\{ism2016004,pse2017002,rsi2017502,sonali\}@iiita.ac.in}}

\maketitle              

\begin{abstract}
With the advancement in the technology sector spanning over every field, a huge influx of information is inevitable. Among all the opportunities that the advancements in the technology have brought, one of them is to propose efficient solutions for data retrieval. This means that from an enormous pile of data, the retrieval methods should allow the users to fetch the relevant and recent data over time. In the field of entertainment and e-commerce, recommender systems have been functioning to provide the aforementioned. Employing the same systems in the medical domain could definitely prove to be useful in variety of ways. Following this context, the goal of this paper is to propose collaborative filtering based recommender system in the healthcare sector to recommend remedies based on the symptoms experienced by the patients. Furthermore, a new dataset is developed consisting of remedies concerning various diseases to address the limited availability of the data. The proposed recommender system accepts the prognostic markers of a patient as the input and generates the best remedy course. With several experimental trials, the proposed model achieved promising results in recommending the possible remedy for given prognostic markers.  

\end{abstract}
\section{Introduction}
Recently, recommender systems have become an important part of many different sectors. Major e-commerce platforms employ the use of recommender systems to display the filtered results for every customer. These recommendations keep updating over time with the aim to improve the user's experience of the platform. The other factors that a recommendation system~\cite{geoviz,tdsNeerja} brings to a platform are: sales boost, enhanced customer engagement, transform shoppers to clients, increase average order value, lower manual work and overhead, and bring more traffic on the e-commerce platform. Recommender systems are also utilized in the following areas: entertainment and media (movie/song/book/news), economy (stocks), banking, telecommunications, etc.

The recommender systems are also termed as SaaS (Software as a Service) and hence could be employed in various other fields given the right dataset. Following this, many recommender system based approaches have been utilized in the healthcare domain to aim for better healthcare services. One of the applications is to recommend remedies to a patient that would be most effective in the treatment process. The system proposed in the paper analyzes the prognostic markers for the patient in question by using the collaborative filtering to match the profiles of other patients that had similar prognostic markers. The top remedies will be listed as the output which have proven to be effective on the other patients’ health status when their prognostic markers were on similar levels.

\begin{figure}[b!]
            \centering
            \includegraphics[width=\textwidth]{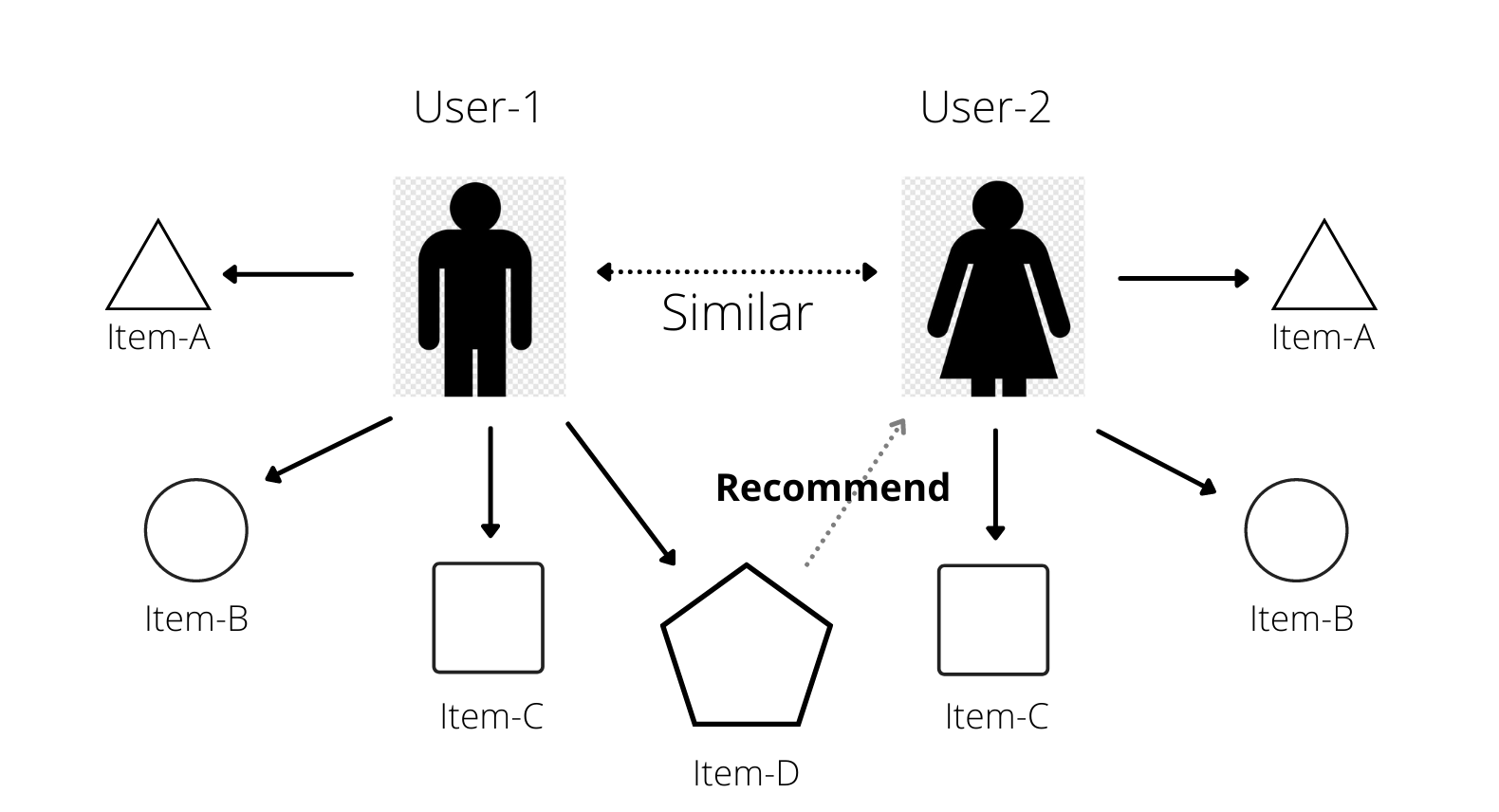}
            \caption{A typical scenario of a collaborative recommender system.}
            \label{fig:rs}
    \end{figure}

\subsection{Motivation and contribution}
Recommender systems are being widely used to recommend the relevant items in the context of e-commerce and infotainment. Recommender systems have helped the businesses because they~\cite{recommenderSystemEmerj} improve the inventory value, user experience etc.

The crux of what is happening behind the scenes is, users are being delivered the relevant content from a huge stack of information. Consequently, when the same approach was applied in the healthcare domain, it gave rise to health recommender systems (HRS). Following this, there was an undeniable boost in automated healthcare and tele-medicine. Health recommender systems (HRS) have been put to use in the following forms:
\begin{enumerate}
    \item Enterprise Resource Planning(ERP system)~\cite{eSezgin}
    \item A Doctor recommendation algorithm~\cite{yfHuang}
    \item Web-based RS suggesting online health resources~\cite{hSchafer}
    \item A diet recommendation system to a patient~\cite{jhKim}
    \item Chronic Disease Diagnosis Prediction and Recommendation System~\cite{asHussein}
\end{enumerate}

According to Park et al.~\cite{park}, the research in the field of health recommender systems has increased immensely but the practical implementations of such systems still requires more research. The major takeaway is that, even with the adequate knowledge in the field of health recommender systems, they are not being put in practical use on a large scale. In this paper, we propose a system in the field of diagnosing and treating diseases in the essence of automating healthcare. The major contributions of the present article are as follows:
\begin{enumerate}
    \item Developing a recommender system that when given a set of symptoms, will perform a diagnosis and then recommend the next best course of treatment.
    \item Creating a data set that contains 'course of treatment' corresponding to their diseases. And these diseases range from mild and acute to chronic states.
    \item Provide possible future improvements for recommender systems in the healthcare domain.
\end{enumerate}

\subsection{Terminologies used}
\begin{enumerate}
    \item \textit{Recommender system (RS)}: recommender systems~\cite{recommenderSystem} are prediction algorithms that are aimed at recommending relevant items or information to the users (usually based on their past preferences). 
    
    \item \textit{Collaborative recommender systems}: These systems recognize the similarities among the given set of users based on their common reviews and provide updated recommendations based on the similar/contrasting features among the users. Fig.~\ref{fig:rs} represents a typical scenario where 'User-1' and 'User-2' have liking for 'Item-A', 'Item-B' and 'Item-C'. This makes both the users similar in the context of ‘the user preference profile’. Hence when 'User-2' adds a new item labeled 'Item-C', the same is recommended to 'User-1' because it’s highly probable that 'User-1' will also prefer 'Item-C' in the future (as both users have preferred the same things in the past).
    
    \item \textit{Bio markers}: A bio marker~\cite{biomarker}, or biological marker is a measurable indicator of some biological state or condition.
    
    \item \textit{Prognostic markers}: Bio markers which can be used to estimate the progression of a medical condition (usually a disease) in an objective manner are called prognostic markers~\cite{prognosisMarker,prognosisMarkerJeffery2018,prognosisMarkerNature}. Prognostic markers are primarily used to divide the patients into categories, aimed at providing precise medicine discovery.
    
    \item \textit{Health recommender systems (HRS)}: recommender systems when applied in the healthcare domain are called health recommender systems.
\end{enumerate}

\subsection{Organization of the paper}
The rest of the paper is divided into several sections. Second section mentions the related works in the field of health recommender systems and the associated challenges. Third section presents the methodology behind building the proposed system. It also mentions the pre-processing of the dataset, and the inclusion of a newly created data file in the dataset. Fourth section presents the results obtained by the proposed approach. Last section concludes the paper with the outlines of the future scope of recommender systems in the healthcare domain.

\section{Related work}
The massive growth and advancements in deep learning algorithms across vivid domains such as healthcare, image processing, etc.~\cite{punn2020multi, agarwal2020unleashing, lu2015recommender, sumanth2020enhanced} has resulted in immense applications by developing real-world applications. In the earlier survey of recommender systems by Park et al.~\cite{park} it was observed that scholastic investigation on recommender systems have expanded fundamentally over the last ten years, but more insights are needed to develop real-world applications. The research field on health recommender systems is potentially wide, however there are less developments in the practical scenarios. In like manner, the current articles on recommender systems should also be reviewed up to the coming age of health recommender systems. Hors-Fraile et al.~\cite{sHorsFraile} also discovered the need of increasing and improving the research in the area of HRS that covers the proposed multidisciplinary taxonomy. This includes the features like integration with electronic health records, incorporation of health promotion theoretical factors and behavior change theories. Kamran and Javed~\cite{mKamran} presented a survey of RS in healthcare and also proposed a hybrid recommender system that takes into account the features like hospital quality (measured objectively based on doctors communication, nurses communication, staff behavior, pain control procedures, medicine explanation, guidance during recovery at home, surrounding cleanliness, quietness in patient’s surrounding) and patients similarity. This recommender system suggests the best hospitals for a patient based on the above factors. 

Pincay et al.~\cite{jPincay} presented a state-of-the-art review providing insights about methods and techniques used in the design and development of HRS(s), focusing on the areas or types of the recommendations these systems provide and the data representations that are employed to build a knowledge base. Sezgin et al.~\cite{eSezgin} outlined the major approaches of HRS which included current developments in the market, challenges, and opportunities regarding HRS and emerging approaches. Huang et al.~\cite{yfHuang} proposed an algorithm which improves the performance of the medical appointment procedure. This algorithm creates a `doctor performance model' based on the reception and appointment status. It also creates a `patient preference model' based on the current and historical reservation choices which help in the accurate recommendation. It prevents the situation where a doctor is under-appointed or over-appointed and the patients are not being treated even if doctors are available. Peito~\cite{jPeito} proposed a HRS for patient-doctor matchmaking based on patients’ individual health profiles and consultation history. Another utility HRS was proposed by Kim et al.~\cite{jhKim} that personalized diet recommended service through considering the real-time vital sign, family history, food preference, and intake of users to solve the limitations in the existing diet recommendation services. 

Hussein et al.~\cite{asHussein} proposed a HRS with the hypothesis that, if a patient's chronic disease diagnosis and set of medical advice are predicted and recommended with high accuracy, it is expected to reflect the improvement of patients' health conditions and lifestyle adjustments along with reducing the healthcare services costs. This can be considered as a `core health recommendation system', as it directly focuses on the disease and the preventive side of the healthcare field, whereas the other HRS usually help a medical institution function better in other aspects. In another similar work, Kuanr et al.~\cite{mKuanr} proposed a HRS to help women by providing information on the features responsible for prognosis of cervical cancer in women. Cheung et al.~\cite{kLong} presented another review which outlines that incorporating multiple filtering, i.e. making a hybrid system could potentially add value to traditional tailoring with regard to enhancing the user experience. This study illustrates how recommender systems, especially hybrid programs, may have the potential to bring tailored digital health forward.

Considering the nature of recommender systems, it's not easy to confine them to some specific sectors. Traditional recommender systems are either collaborative or content-based (broadly speaking). In HRS, which type of recommender system should be used depends on the application. For instance, collaborative filtering might be used in an educational context, whereas content-based filtering would prove to have more impact in creating a doctor recommendation algorithm that takes into account the performance of doctors as well. In collaborative filtering, only the objective information regarding the items are stored, whereas, in content-based filtering, more comprehensive information is stored which gives rise to the following major privacy issue~\cite{eSezgin}: `Combining data from multiple users (probably from different geographical locations) can be seen as an intrusion to the individual private data. It may even uncover some confidential data of healthcare institutions'. This poses a major challenge that violates the delicate topic about privacy which must be confidential in a healthcare system.

Apart from this major healthcare sector confined flaw, in general there are some other basic challenges that a recommender system faces (which are also applicable to HRS):

\begin{enumerate}
    \item \textit{Data sparsity}: If the recommender system is employed in very few places then the performance of the system may not be very promising. As it will only infer suggestions based on the data samples considered within its limited range, it will not follow into the standard footsteps of a recommender system which usually works on large data samples.
    
    \item \textit{Scalability}: If the users in the system scale to a very large number, say in millions, then the collaborative filtering algorithm will fail. The linear run time algorithm (i.e. $\mathcal{O}(n)$ time-complexity) is undesirable in healthcare scenarios because the results should be generated in real-time.
    
    \item \textit{Diversity and the long tail}: The system won’t be able to recommend remedies with a limited amount of historical data. If a remedy is not recommended to a set of patients with similar prognostic markers as the current user, then that remedy is unlikely to be recommended to the current user, even though that specific remedy could prove more beneficial than the rest.
\end{enumerate}

In the healthcare sector, every peripheral context is as important as working on ailments and their remedies. All five health recommender systems~\cite{eSezgin,yfHuang,jhKim,nMohammadi} mentioned in Section 1.1, are aimed at providing recommendations in the healthcare context. Among these HRS(s), Hussein et al.~\cite{asHussein} approach works on predicting diseases or recommending treatment. The CDD recommender system (as they coined it), acts as a core HRS since it predicts the disease and recommends medical advice. This type of HRS acts as an extra tool, by assisting the physicians and patients in controlling and managing the diseases. They have employed `decision tree' algorithm in `random forest' manner for disease prediction and used a `unified collaborative filtering' approach for advice recommendation. This complete model seems a breakthrough in the HRS sector, however the model is built to predict and diagnose only the `chronic diseases'. Following this context, the present research work aims to develop an HRS that:

\begin{enumerate}
    \item Acts as a core HRS (acting as a tool to help recommend medical remedies).
    \item Provides medical remedies to a wide range of diseases, not just the chronic diseases.
\end{enumerate}

\section{Methodology}
The recommender systems are usually of the following types:
\begin{enumerate}
    \item Collaborative filtering: It works by locating peer items with a rating history similar to the current item, and then these nearest neighbours are used to generate recommendations.

    \item Content-based filtering: It matches the new items for the user with the items previously rated by the same user.

    \item Multi-criteria recommender systems: It takes into consideration multiple criterion for suggesting an item to a user.

    \item Mobile recommender systems: It aims to provide personalised recommendations while keeping in mind that the mobile data is more complex than the regular data and the protection of privacy needs to be incorporated carefully.

    \item Hybrid recommender systems: It works by combining collaborative filtering, content-based filtering and other approaches into one system.

    \item Session-based recommender systems: It doesn’t take into account the older history of the user but the usage patterns of current session only.
\end{enumerate}

As mentioned earlier, collaborative filtering and content-based filtering are the most common approaches to build a recommendation system. The proposed model of HRS is built using the collaborative filtering technique.

\subsection{Collaborative filtering}
Collaborative filtering~\cite{collaborativeFiltering} as the name suggests, employs the use of collaboration. The underlying presumption of this approach is that if an individual $A$ has a similar assessment as an individual $B$ on a context, $A$ is probable to have $B$'s assessment on a different issue in comparison to that of a randomly picked individual. For instance, a collaborative filtering recommendation system for shopping preferences could make predictions about which outfits and accessories a client would like, given an partial list of that client's preferences. These preferences may include likes or dislikes, frequency of buying from a particular brand, the average spending amount, etc. It should be noted that, even though these predictions use data gathered from numerous clients, ultimately provides tailored predictions to individual client(s). This contrasts from the simpler methodology of giving a normalized rating for every item of interest.

The analogy to the system proposed in this paper holds as: If a person $A$ was cured by the same treatment as person $B$ given the same set of symptoms, $A$ and $B$ are more likely to be cured by the same treatment for a new set of common symptoms.

\subsection{Dataset synthesis}
Acquiring the dataset for this system was one of the biggest challenges. There are plenty of datasets publicly available for the healthcare domain but most of them conform only to some specific category of illness (like heart diseases only, nervous system disorder only, etc). This system requires a dataset that contains a list of diseases spread over various domains. The base dataset\footnote{\url{https://www.kaggle.com/plarmuseau/sdsort}\label{dataset}} hence chosen is taken from the profile of P. Larmuseau, Pharmacist at Gaver Apotheek (Harelbeke, Flanders, Belgium). The dataset consists of 8 files in $.csv$ format. The primary data from the dataset contained the information arranged in tuples. Following list shows the labeling (of the tuples) as found in the files of the dataset in the format as \textit{(file name: (tuple labels) = (corresponding alias))}:

\begin{enumerate}
    \item sym\_t.csv : (syd, symptom) = (Symptom identifier, Symptom name)
    \item dia\_t.csv : (did, diagnose) = (Disease identifier, Disease name)
    \item diffsydiw.csv : (syd, did, wei) = (Symptom identifier, Disease identifier, Weight of the symptom on the disease)
    \item prec\_t.csv : (did, diagnose, pid) = (Disease identifier, Disease name, treatment course)
\end{enumerate}

\begin{figure}[h]
        \centering
        \includegraphics[width=0.7\textwidth]{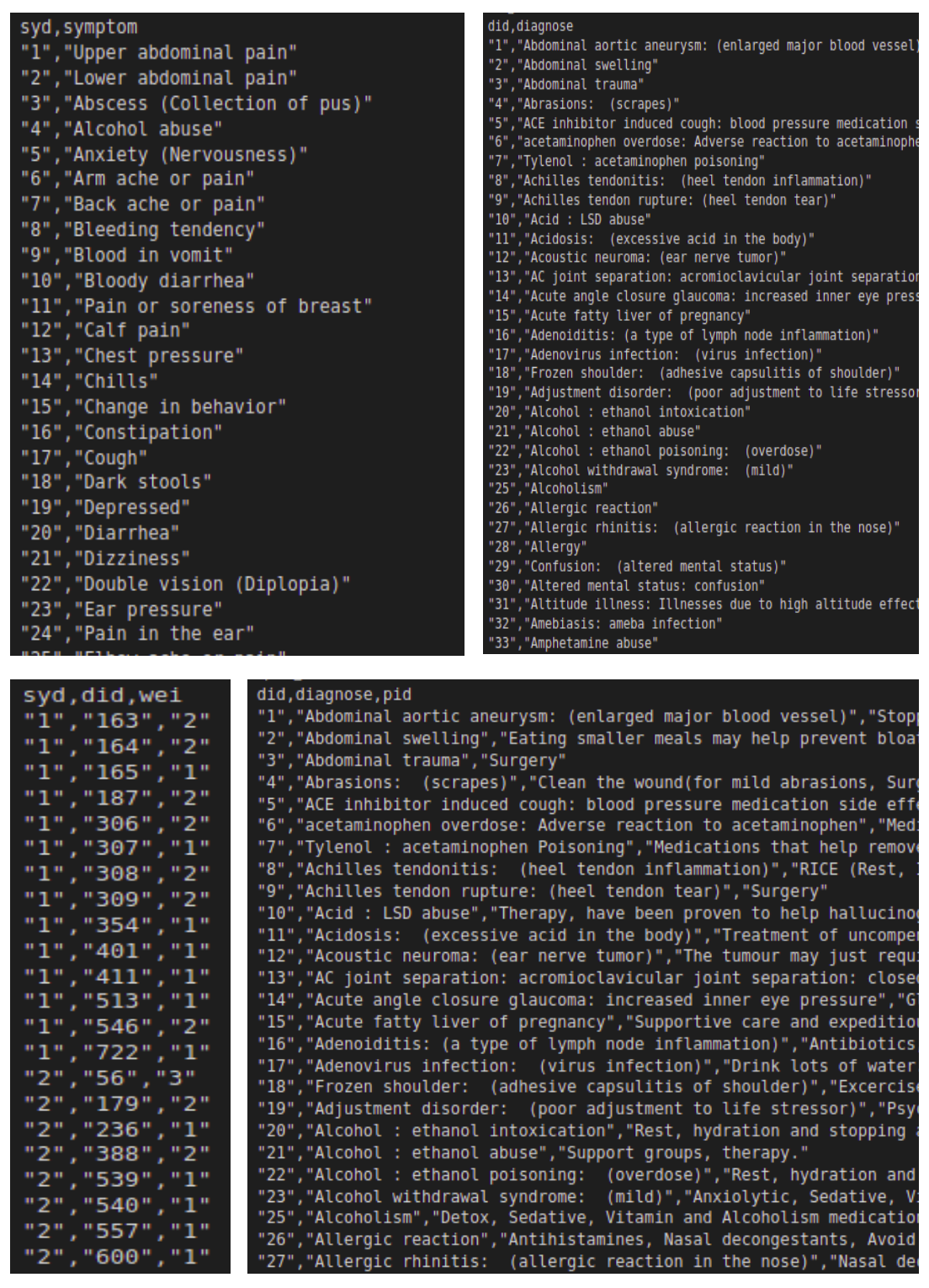}
        \caption{The first three files are from the base dataset. The fourth file with column labels as (Disease Identifier, Disease name, treatment course) is synthesized for the proposed remedy recommendation system. }
        \label{fig:dataset}
\end{figure}

A total of 1,167 diseases and 273 symptoms are listed in the dataset. Fig.\ref{fig:dataset} shows the snapshots for the above mentioned files in order.

There was no good data sources available for all diseases (from generic to chronic) and their treatment courses. Hence the data file $prec\_t.csv$\footnote{\url{https://github.com/sud0lancer/Diagonosis-Precaution-dataset}\label{precautionRepo}} (name as in the code repository) is created by exploring several medical websites and resources~\cite{webmd,healthline,medlineplus,medicalnewstoday,msdmanuals,medscape,cleavelandclinic,mayoclinic,albertEinsteinHospital,apolloHospitals,collegeOfMedicineUniversityofIbadan,columbiaAsiaHospitals,hms,Hoken,cancerGov,dermnetnz,nih,urology,drugabuse}.

\subsection{Dataset pre-processing}
In order to develop the complete dataset for the proposed system, the base dataset\textsuperscript{\ref{dataset}} was cleaned using the following steps:
\begin{enumerate}
    \item Dropping the rows if any of the attributes was $NULL$.
    \item Removing unrecognizable delimiters and replacing them with commas(,).
\end{enumerate}

\subsection{Building a sparse matrix}
The pre-processed data from the data-files is used to create a new matrix which is sparse in nature. For instance, the columns from files: sym\_t.csv - (syd, symptom), dia\_t.csv - (did, diagnose) and diffsydiw.csv - (syd, did, wei) are transformed into a sparse matrix $Data(i,j)$ such that $Data[i][j] \geq 0$ represents the weight of the $j^{th}$ symptom on $i^{th}$ disease, where higher value represents larger weight of a symptom for a disease and $0$ represents that the symptom doesn’t give rise to the corresponding disease. This matrix is considered as the source from where the system will generate the recommendation.

\subsection{Normalization using BM25 weighting}
Normalization was done to calculate the average weight (or importance) of a symptom for a disease in presence of other symptoms. BM25 weighting~\cite{bm25tfidf} scheme is used for this purpose. BM25 is considered to be a better version of the TF-IDF ranking functions. Main motive of these functions is to estimate the relevance (score, ranking) of a term in a huge text corpus. These functions employ the frequency and rarity of a term to compute their importance to the corpus. TF-IDF uses the Eq.~\ref{eq1} to compute the relevance score:

\begin{equation}
    R_{score}(D, T) = termFrequency(D, T) * log\left(\frac{N}{docFrequency(T)}\right)
    \label{eq1}
\end{equation}
Here:
\begin{enumerate}
    \item $R_{score}(D, T)$ = score of a term $T$ in a document $D$,
    \item $termFrequency(D, T)$ = how many times does the term $T$ occur in document $D$.
    \item $docFrequency(T)$ = in how many documents does the term $T$ occur.
    \item $N$ = size of the search index or corpus.
\end{enumerate}

Scores for all such documents in the corpus are added to get the final score (ranking) of a term. In contrast, BM25 adds modifications to compute the ranking score as follows:

\begin{enumerate}
    \item A document length variable is added, where larger documents are penalised having the same term frequency as those of smaller documents. 
    \item After the term’s frequency has reached saturation, further occurrences don’t affect its score.
\end{enumerate}

In BM25~\cite{bm25tfidf}, the ranking score can be computed using Eq.~\ref{eq2}.

\begin{equation}
    R_{score}(D, Q) =
    \sum_{t \in Q}^{} \frac{f_{t,D} \cdot (k_1 + 1)}{f_{t,D} + k_1 \cdot (1 - b + b \cdot \frac{|D|}{avg(dl)}) } 
    \cdot \log \frac{N - n_t + 0.5}{n_t + 0.5} 
    \label{eq2}
\end{equation}

Here:
\begin{enumerate}
    \item $\sum_{t \in Q}^{}$ = sum the scores of each query term,
    
    \item $\frac{f_{t,D}}{f_{t,D} + k_1 }$ = term frequency saturation trick,
    
    \item $\frac{1}{(1 - b + b \cdot \frac{|D|}{avg(dl)})}$ = adjust saturation curve based on document length,
    
    \item $\frac{N - n_t + 0.5}{n_t + 0.5}$ = probabilistic flavour of IDF.
\end{enumerate}

In this case, the analogy holds as: `symptom’ is `term’, `list of diseases’ is `the huge text corpus’. After the normalization, the sparse matrix will now contain updated values. For example, $Data[i][j] = 3$ will have changed to either $2.8736$ or $3.1252$ (exact values may vary) depending on the effect of the other symptoms on the corresponding disease.

\subsection{SVD and cosine similarity}
Single value decomposition (SVD) in the context of recommendation systems is used as a collaborative filtering (CF) algorithm. It is used as a tool to factorize the sparse matrix to get better recommendations.

SVD is a matrix factorization technique that splits a matrix into products of two or more matrices such that when these constituent matrices are multiplied back, they will return the original matrix. SVD is a most common method for dimensionality reduction. This implies that when the data is available in many dimensions (i.e. there are a lot of attributes projected in same (or different) directions), it is not easy to infer information due to the curse of dimensionality, however, if the data is reduced to less dimensions, then it would be easier to visualize the data and extract the desired information.

Let $R \in \mathbb{R}^{m \times n}$ be the original data matrix. Then after applying SVD, $R$ breaks into the following 3 matrices as shown in Eq.~\ref{eq3}:
\begin{enumerate}
    \item $U$ is a $m \times r$ orthogonal left singular matrix,
    \item $V$ is a $r \times n$ orthogonal right singular matrix,
    \item $S$ is a $r \times r$ diagonal matrix, such that
\end{enumerate}

\begin{equation}
R = USV^T
\label{eq3}
\end{equation}

The SVD decreases the dimensions of the original matrix $R$ from $m \times n$ to $m \times r$ and $r \times n$ by extracting its latent factors. In our case, $R \in \mathbb{R}^{1145 \times 272}$ is reduced as: $U \in \mathbb{R}^{1145 \times 50}$, $S \in \mathbb{R}^{50 \times 50}$ and $V \in \mathbb{R}^{50 \times 272}$. The matrices $U$ and $V$ are used to find the recommendations.

Cosine similarity is the measure of similarity between two vectors. This similarity is calculated by measuring the cosine of the angle between the two vectors which may be projected into multidimensional space. It can be applied to items available in a dataset to compute similarity among themselves. Similarity between two vectors ($A$ and $B$) is calculated by dividing the dot product of the two vectors by the magnitude value as shown in Eq.~\ref{eq4}. It is worth noting that the similarity (or CS score) of the given two vectors is directly proportional to the angle between them. 

\begin{equation}
    \cos\theta = \frac{A \cdot B}{||A|| \cdot ||B||} =
    \frac{\sum_{i = 1}^{n} Ai \cdot Bi}{\sqrt{\sum_{i = 1}^{n} (Ai)^2} \cdot \sqrt{\sum_{i = 1}^{n} (Bi)^2}}
    \label{eq4}
\end{equation}

In the proposed system, cosine similarity finds the $n$ rows (where each row is represented by a disease) from the decomposed matrices, that have the maximum sum of the symptoms weight(which in turn means the diseases having the maximum matching symptoms). Here $n$ is the number (a manual threshold) of diseases that we wish to generate for the given inputs. 

\section{Results and discussion}
The proposed system was tested in two phases in order to determine the working  and the performance of the system respectively. In the first phase, the experiments are conducted keeping in mind the related and unrelated symptoms experienced by the patients that signify the real life scenarios. Sometimes a patient might be experiencing multiple symptoms but most of them hint towards a common disease and in the other cases, those symptoms may be completely unrelated to each other and hence the patient might be having multiple diseases. In the case of related symptoms, the HRS is expected to recommend the remedy for the disease that is most likely to happen because of the given multiple symptoms that are related to each other. And in the case of unrelated symptoms, the HRS must recommend the remedies for all the different possible diseases. Table.\ref{tab1} shows the two cases as mentioned above. `Case:1 Unrelated-symptoms' has an array of `symptom\_id' as input. It predicts the probable disease(s) and then recommends the best treatment. Likewise for `Case2: Related-symptoms'.

\begin{table}[]
\caption{The remedy recommendation results of the proposed HRS for the given symptoms.}
\label{tab1}
\begin{tabular}{|p{0.6in}|p{0.62in}|p{0.81in}|p{1.25in}|p{1.25in}|}
\hline
\textbf{Case} & \textbf{Symptom ID} & \textbf{Symptoms} & \textbf{Most probable   disease(s)} & \textbf{Best treatment(s)} \\ \hline
\multirow{5}{0.6in}{Related symptoms} & 1          & Upper   abdominal pain & \multirow{5}{1.25in}{1: Ventral hernia: bulging of the abdominal wall. \\ 2: Diverticulosis: weakening of the large intestine wall.} & \multirow{4}{1.25in}{1: Eating   smaller meals may help prevent bloating and swelling.} \\ \cline{2-3}
& 2 & Lower abdominal pain & & \\ 
& & & & \\
& & & & \\
\hline
\multirow{6}{0.6in}{Unrelated symptoms}   & 2          & Lower abdominal pain & \multirow{6}{1.25in}{1: Ventral hernia: bulging of the abdominal wall.\\ 2: Vitiligo: loss of skin pigment} & \multirow{6}{1.25in}{1: Laparoscopic surgery. \\ 2: Photodynamic therapy, Medications: Steroid and Immunosuppresive drug.} \\ \cline{2-3} 
& 81 & Rash & & \\
& & & & \\
& & & & \\
& & & & \\
& & & & \\
\hline
\end{tabular}
\end{table}

\begin{figure}[h]
        \centering
        \includegraphics[width=1\textwidth]{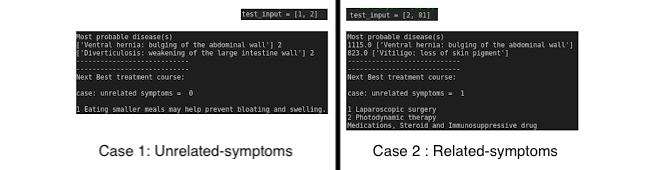}
        \caption{The output shows the recommendations given by the system when given symptom(s) as input.}
        \label{fig:results}
\end{figure}

In both the cases, the system takes in an array of symptom IDs, then it predicts the most probable diseases as a result of the symptoms and then recommends the course of treatment. Fig.\ref{fig:results} shows the output produced by the HRS for both the cases of related and unrelated symptoms. 

The second phase consists of further evaluations which incorporates the analysis of the quality of the predictions based on various types of testing. The first level of testing is the `sanity testing'. In sanity testing, the dataset is divided into two halves while preserving the uniformity of the symptoms subgroups in the dataset. Both the halves are fed as the input to the system, and the corresponding similarity matrices generated must be as close as possible to the original similarity matrix, i.e. it must show minimal difference in the values (along the diagonal) which will indicate that the similarity matrices are very similar. It implies that there exists least dependency on the data and the type of data that is fed to the system. Euclidean distance is used to find the similarity between the matrices. Henceforth it can be said that the proposed system is un-biased towards data. Matrices $M_1$ and $M_2$ show that the euclidean distance matrix has all diagonal values near $0$, implying that the two similarity matrices are similar. $M_1$ is composed of two similarity matrices belonging to full dataset and one of the halves of the dataset, respectively. Similar results hold for the similarity of the other half of the dataset with the full dataset. $M_2$ is composed of two similarity matrices belonging to both the halves of the dataset.

\begin{equation*}
M_1 = 
\begin{bmatrix}
\textbf{0.3610} & 2.8490 & 3.3920 & 4.8063 & ...\\
2.9624 & \textbf{0.3843} & 3.2234 & 3.8602 & ...\\
3.3868 & 3.1321 & \textbf{0.3062} & 3.3776 & ...\\
4.8199 & 3.8164 & 3.3867 & \textbf{0.2363} & ...\\
... & ... & ... & ... & ...
\end{bmatrix}
\end{equation*}

\begin{equation*}
M_2 = 
\begin{bmatrix}
\textbf{0.3031} & 2.8507 & 3.3858 & 4.8042 & ...\\
2.8626 & \textbf{0.1469} & 3.1463 & 3.8111 & ...\\
3.4013 & 3.1316 & \textbf{0.4758} & 3.3871 & ...\\
4.8306 & 3.8184 & 3.3974 & \textbf{0.3423} & ...\\
... & ... & ... & ... & ...
\end{bmatrix}
\end{equation*}

\begin{figure}[h]
        \centering
        \includegraphics[width=1\textwidth]{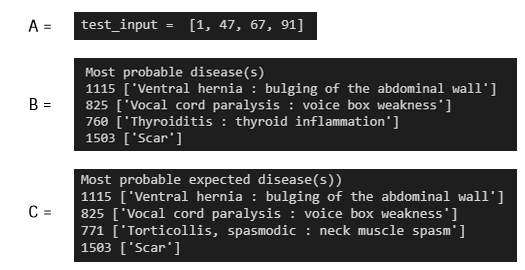}
        \caption{Given a set of symptoms(A), the predicted(B) and expected outputs(C) are shown respectively.}
        \label{fig:test3}
\end{figure}

The second level of testing is the regression testing, in which we use a subset of the training data to generate the output and match with the training set, which proves that the model is correctly created and has proper similarity matrices. Fig.\ref{fig:test3} shows that when given a set of symptoms, system predicts $3$ out of $4$ diseases which resemble the ground truth (since they had the maximum weight in the training dataset). The remaining $4^{th}$ prediction is also correct but it had a lower weight in the training dataset. Hence it can be said that the system predicts the most probable disease(s) for a given set of symptoms together, not the most probable disease for each individual symptom(s).

\section{Conclusion and Future scope}
Indeed recommender systems play a major role in everybody’s daily life covering online shopping, movie streaming, etc. With state-of-the-art potential of recommendar system, this can be extended to healthcare department to aid in a variety of tasks such as managing resources of a healthcare institution, replacement suggestions for equipment(s), recommending medical advice and suggestions, etc. In this paper, the proposed model recommends remedies for the predicted disease(s) followed from the given symptoms by using the generated dataset\footnote{\url{https://github.com/sud0lancer/Diagonosis-Precaution-dataset}} consisting of a list of most favourable remedies corresponding to a wide range of disease(s). The future scope for this system includes improving the dataset by categorising the mentioned remedies under labels such as: self-care, medication, surgical procedures, non-surgical procedures, therapies, etc., incorporating more filtering algorithms for better results, creating a web based UI for better interaction with the proposed system. 

\section*{Acknowledgment}
This research is supported by “ASEAN- India Science \& Technology Development Fund (AISTDF)”, SERB, Sanction letter no. – IMRC/AISTDF/R\&D/P-6/2017. Authors are also thankful to the authorities of “Indian Institute of Information Technology, Allahabad at Prayagraj”, for providing us with the infrastructure and necessary support.

\bibliographystyle{splncs04}
\bibliography{mybibliography}

\end{document}